\documentclass[article, twocolumn,
   aps, pra,
  amsmath,amssymb,
  longbibliography,
  ]{revtex4-1}
\usepackage{graphicx,color}
\usepackage{amsmath}
\usepackage{natbib}
\usepackage{epsfig}
\begin{document}

\title{Excess energy of strongly coupled one-component plasma from variational approach  }

\author{S. A. Khrapak\email{Sergey.Khrapak@gmx.de} 
and A. G. Khrapak
}
\affiliation{Joint Institute for High Temperatures, Russian Academy of Sciences, 125412 Moscow, Russia}

\begin{abstract}
The excess energy of the one-component plasma fluid is calculated using the variational approach combined with different variants of the excess entropy of the hard-sphere fluid, which is used as a reference system. Our comparison with recent Monte Carlo results for the excess energy of the one-component plasma identifies the Percus-Yevick virial entropy as the most accurate entropy to be used in the variational calculation of this kind. The reason for this and potential developments of the present analysis are briefly discussed. We demonstrate that the original Rosenfeld-Tarazons scaling of the thermal component of the excess energy of the one-component plasma fluid is in excellent agreement with recent Monte Carlo results. 
\end{abstract}

\date{\today}

\maketitle

\section{Introduction}

To estimate the properties of a many-body system, a relation can be made to another reference system whose properties are known with better accuracy. In particular, the Bogoliubov variational inequality is widely used for this purpose~\cite{HansenBook}. It represents a fundamental result in statistical physics imposing an upper bound on the Helmholtz free energy of a many-body system:
\begin{equation}\label{Bogoliubov}
F\leq F_0+\langle H-H_0\rangle_0,
\end{equation}
where $F$ and $F_0$ are Helmholtz free energies of the actual and reference systems, while $H$ and $H_0$ are their respective Hamiltonians, evaluated at a given state of the reference system. For classical systems of particles interacting with pairwise potentials, the Hamiltonians differ only by the potential (excess) energies evaluated for a given state of the reference system. This can be expressed through the radial distribution function (RDF) $g_0(r)$ of the reference system by means of the integral energy equation~\cite{HansenBook}. It is particularly tempting to use an assembly of hard spheres (HS) as a reference system because the potential energy is identically zero, the properties of the system, such as $g_0(r)$, are quite well known and depend on a single parameter -- the packing fraction $\eta=\pi n\sigma^3/6$, where $n$ is the density of spheres and $\sigma$ is their diameter. Adopting this choice, we can rewrite the Bogoliubov inequality as
\begin{equation}\label{Bogoliubov1}
F\leq -TS_{0}(\eta) + \frac{N}{2}n\int g_0(r;\eta)\phi(r)d{\bf r},
\end{equation}
where $T$ is the temperature, $S_0$ is the entropy of the hard-sphere system (related to an appropriate equation of state), $N$ is the number of particles, and $\phi(r)$ is the interaction potential of the system considered. The right-hand side (RHS) of equation (\ref{Bogoliubov1}) depends on the packing fraction $\eta$, which acts as a variational parameter. Minimizing the RHS with respect to $\eta$ would produce the best estimate of the free energy of the system considered. This is how the variational calculation with the HS reference system works in practice.

Naturally, it should be expected that the HS reference is more appropriate for systems with steep interactions such as the Lennard-Jones system~\cite{MansooriJCP1969a,MansooriJCP1969b} or inverse power potentials with sufficiently large exponents~\cite{RossPRA1973,YoungJCP1984}. Surprisingly, it remains a useful phenomenological model even in the case of extremely soft and long-range Coulomb-like interactions, although not all of its realizations deliver satisfactory accuracy~\cite{RossPRA1981}. Actually, it turns out that the procedure is rather sensitive to the exact form of the HS entropy. Not the accuracy of the $S_0(\eta)$ term {\it per se}, but rather its consistency with the RDF $g_0(r;\eta)$ used to calculate excess energy determines the accuracy of the variational calculation.    

In this paper, we consider a strongly coupled one-component plasma (OCP), which is a classical fluid of charged particles that interact through a Coulomb interaction potential and are immersed in a uniform compensating background of opposite charge to ensure quasineutrality~\cite{BausPR1980,IchimaruRMP1982}. We calculate the excess internal energy employing a variational approach with the HS fluid as a reference system using five different equations of state (EoS) of the HS fluid. In this way, we quantify the accuracy of different approximations for $S_0(\eta)$ in this important special case. Importantly, we base our comparison on results from recently reported molecular dynamics (MD) simulations of the thermodynamical properties of HS fluids~\cite{Pieprzyk2019} and Monte Carlo (MC) simulations of OCP fluids using the angular-averaged Ewald potential (AAEP)~\cite{DemyanovPRE2022,DemyanovArxiv2025}. Since these results seem to reach the level of accuracy needed for most practical applications, a systematic evaluation of the impact of different assumptions for HS entropy appears to be both relevant and timely. In addition, we use this opportunity to update the applicability of the Rosenfeld-Tarazona scaling in the case of the OCP fluid.  

\section{Calculation}

The OCP is characterized by a single dimensionless Coulomb coupling parameter $\Gamma = q^2/aT$, where $q$ is the electric charge and $a=(4\pi n/3)^{-1/3}$ is the Wigner-Seitz radius. At $\Gamma\ll 1$ the weakly coupled regime is realized and the OCP exhibits a gas-like behavior. At $\Gamma\gtrsim 1$ strong coupling takes place and the OCP exhibits fluid-like behavior. In fact, the center of the crossover from gas-like to liquid-like dynamics has been recently identified at $\Gamma\sim 10$~\cite{HuangPRR2023,KhrapakPoF2022,KhrapakPoP2025}. At $\Gamma\gtrsim 50$ vibrational dominance governs the dynamical behavior and transport properties of the OCP~\cite{KhrapakMolecules12_2021,KhrapakPhysRep2024}. The fluid-solid phase transition occurs at $\Gamma\simeq 174$, where the Helmholtz free energy of the fluid and solid phases intersect and the OCP crystallizes into a body-centered cubic lattice~\cite{IchimaruRMP1982,DubinRMP1999,KhrapakCPP2016}. The existence of a glass transition at even high $\Gamma$ has also been discussed in the literature from different perspectives~\cite{TanakaPRA1987,IchimaruPRL1986,CardenasPhysB2004,YazdiPRE2014,KhrapakJCP2024_glass}. Here, our main interest is the strongly coupled fluid regime. 

Omitting the ideal gas terms, we rewrite Eq.~(\ref{Bogoliubov1}) in conventional reduced excess units~\cite{BausPR1980}
\begin{equation}\label{Bogoliubov2}
 f_{\rm ex}\leq -s_{\rm ex}(\eta)+u_{\rm ex}(\Gamma,\eta).   
\end{equation}
The calculation of excess energy should be adapted to reflect the particular properties of the OCP. Taking into account the presence of neutralizing background, the Coulomb interactions potential $\phi(r)=q^2/r$, and introducing the reduced distance $x=r/\sigma$, the excess energy term becomes
\begin{equation}
u_{\rm ex}(\Gamma,\eta)=6\Gamma\eta^{2/3}\int_0^{\infty}\left[g_0(x;\eta)-1\right]xdx.    
\end{equation}
The integral can be calculated by noting that
\begin{multline}
\int_0^{\infty}\left[g_0(x;\eta)-1\right]xdx  = \\ =\lim_{t\rightarrow 0}\int_0^{\infty}e^{-tx}\left[g_0(x;\eta)-1\right]xdx=\lim_{t\rightarrow 0}\left[G(t,\eta)-\frac{1}{t^2}\right].    
\end{multline}
In Percus-Yevick (PY) theory, the function $G(t,\eta)$ is known analytically~\cite{Wertheim1963}. Evaluating the limit is a nice exercise yielding
\begin{equation}\label{uex}
u_{\rm ex}(\Gamma, \eta)= -3\Gamma\eta^{2/3}\frac{1-\tfrac{1}{5}\eta+\tfrac{1}{10}\eta^2}{1+2\eta}.    
\end{equation}
This result was derived by Jones~\cite{JonesJCP1971}, although with a misprint in the original version~\cite{StroudPRA1976}. 

The excess entropy is related to the compressibility factor $Z=P/nT$ (we measure the temperature in energy units so that $Z$ is a dimensionless quantity) through an integral equation
\begin{equation}\label{sex}
  s_{\rm ex}(\eta)=-\int_0^\eta\frac{Z(\eta')-1}{\eta'}d\eta'.
\end{equation}

In the following, we test five different EoS for the HS fluid. The first two are the conventional PY results~\cite{Wertheim1963,Thiele1963} obtained either through the {\it virial} (pressure) route
\begin{equation}\label{Zv1}
  Z_{\rm v}(\eta)=\frac{1+2\eta+3\eta^2}{(1-\eta)^2},
\end{equation}
or through the {\it compressibility} route
\begin{equation}\label{Zc1}
  Z_{\rm c}(\eta)=\frac{1+\eta+\eta^2}{(1-\eta)^3}.
\end{equation}
The difference is due to thermodynamic inconsistency related to the approximate character of the PY theory. A more accurate {\it Carnahan-Starling} EoS~\cite{CarnahanJCP1969} can be regarded as a linear superposition of $Z_{\rm v}$ and $Z_{\rm c}$,
\begin{equation}\label{ZCS1}
  Z_{\text{CS}}(\eta)=\frac{1+\eta+\eta^2-\eta^3}{(1-\eta)^3}.
\end{equation}
The fourth EoS is a more recent result, which combines the PY theory with the {\it chemical potential} route~\cite{SantosPRL2012}  
\begin{equation}\label{Zmu1}
  Z_{\mu}=-9\frac{\ln(1-\eta)}{\eta}-8\frac{1-\frac{31}{16}\eta}{(1-\eta)^2}.
\end{equation}
The fifth is a semi-empirical {\it modified Kolafa, Labik and Malijevsky} (mKLM) EoS based on the functional form suggested in Ref.~\cite{KolafaPCCP2004} and modified in view of the new MD simulation results~\cite{Pieprzyk2019}. This EoS has the following analytical form
\begin{equation}
Z_{\rm mKLM}(\eta)=1+\sum_{j=1}^{8}B_i x^j+B_{10}x^{10}+B_{14}x^{14}+B_{22}x^{22},   
\end{equation}
where $x=\eta/(1-\eta)$ and the coefficients are $B_1=4$, $B_2=6$, $B_3=2.364768$, $B_4=-0.8698551$, $B_5=1.1062803$, $B_6=-1.1014221$, $B_7=0.66605866$, $B_8=-0.03633431$, $B_{10}=-0.20965164$, $B_{14}=0.10555569$, and $B_{22}=-0.00872380$. It gives an excellent representation of the thermodynamics of the HS fluid (stable and metastable) up to $\eta\simeq 0.534$~\cite{Pieprzyk2019}.

We do not derive the corresponding expressions for $s_{\rm ex}$ because these are not required in the variational calculation. Since we need to minimize the RHS of Eq.~(\ref{Bogoliubov2}), the derivatives of the excess energy and of the entropy with respect to $\eta$ are of primary interest. We get 
\begin{equation}\label{du}
 \frac{\partial u_{\rm ex}}{\partial\eta}=-\Gamma\frac{(1-\eta)^2(2+\eta)}{\eta^{1/3}(1+2\eta)^2}
\end{equation}
and 
\begin{equation}\label{ds}
 \frac{\partial s_{\rm ex}}{\partial\eta}=-\frac{Z(\eta)-1}{\eta}.
\end{equation}
Requiring the derivative of the RHS of Eq.~(\ref{Bogoliubov2}) to be zero, we obtain a simple relation between $\Gamma$ and $\eta$
\begin{equation}\label{Gamma}
\Gamma = \frac{\left[Z(\eta)-1\right](1+2\eta)^2}{\eta^{2/3}(1-\eta)^2(2+\eta)}.
\end{equation}
This relation obviously depends on the particular form of the HS EoS chosen for calculation. The excess energy, as a function of $\Gamma$, can then be readily obtained from Eq.~(\ref{uex}). This provides easy access to the thermodynamics of the OCP fluid in a simple parametric form.

\section{Results}

Among the five HS EoS considered, it can be expected that the mKLM EoS is the most accurate among the currently available, as Fig.~1 from Ref.~\cite{Pieprzyk2019} demonstrates. All EoS are relatively accurate in the low-density domain (small $\eta$) as the two first virial coefficients are exact. The deviations become pronounceable as $\eta$ increases. To demonstrate the extent of these deviations, we have calculated several important properties of the HS fluid at the packing fraction corresponding to solidification at $\eta\simeq 0.4918$~\cite{Pieprzyk2019}. Three properties have been selected, the compressibility factor itself, the excess entropy, and the sound velocity. The results are summarized in Tab.~\ref{Tab1}.       

\begin{table}
\caption{\label{Tab1} Compressibility factor $Z$, reduced excess entropy $s_{\rm ex}$, and the reduced sound velocity $c_s/v_{\rm T}$ (where $v_{\rm T}$ is the thermal velocity) of the HS fluid at freezing, $\eta=0.4918$~\cite{Pieprzyk2019}. Calculations are performed using five EoS models: PY through the virial route (PY-v), PY through compressibility route (PY-c), PY through chemical potential route (PY-{$\mu$}), Carnahan-Starling EoS (CS), and modified Kolafa, Labik and Malijevsky EoS (mKLM). }
\begin{ruledtabular}
\begin{tabular}{crrrrr}
  & PY-v & PY-c & PY-{$\mu$} & CS & mKLM   \\ \hline
$Z$  & 10.49 & 13.21 & 10.92 & 12.30 & 12.31 \\
$s_{\rm ex}$ & -4.45 & -4.98 & -4.55 & -4.81 & -4.82  \\
$c_{\rm s}/v_{\rm T}$  & 10.66 & 13.23  & 11.07 & 12.39 & 12.40  \\
\end{tabular}
\end{ruledtabular}
\end{table}

We observe that the CS EoS delivers results that are quite close to the most accurate mKLM. Other approaches lead to considerable deviations. The relative magnitude of these derivations can be as high as $\sim 20\%$. The PV-v and PV-$\mu$ approaches underestimate the compressibility factor and the sound velocity and overestimate the excess entropy. In contrast, the PY-c approach overestimates the compressibility factor and the sound velocity and underestimates the excess entropy.  

\begin{figure}
\includegraphics[width=8cm]{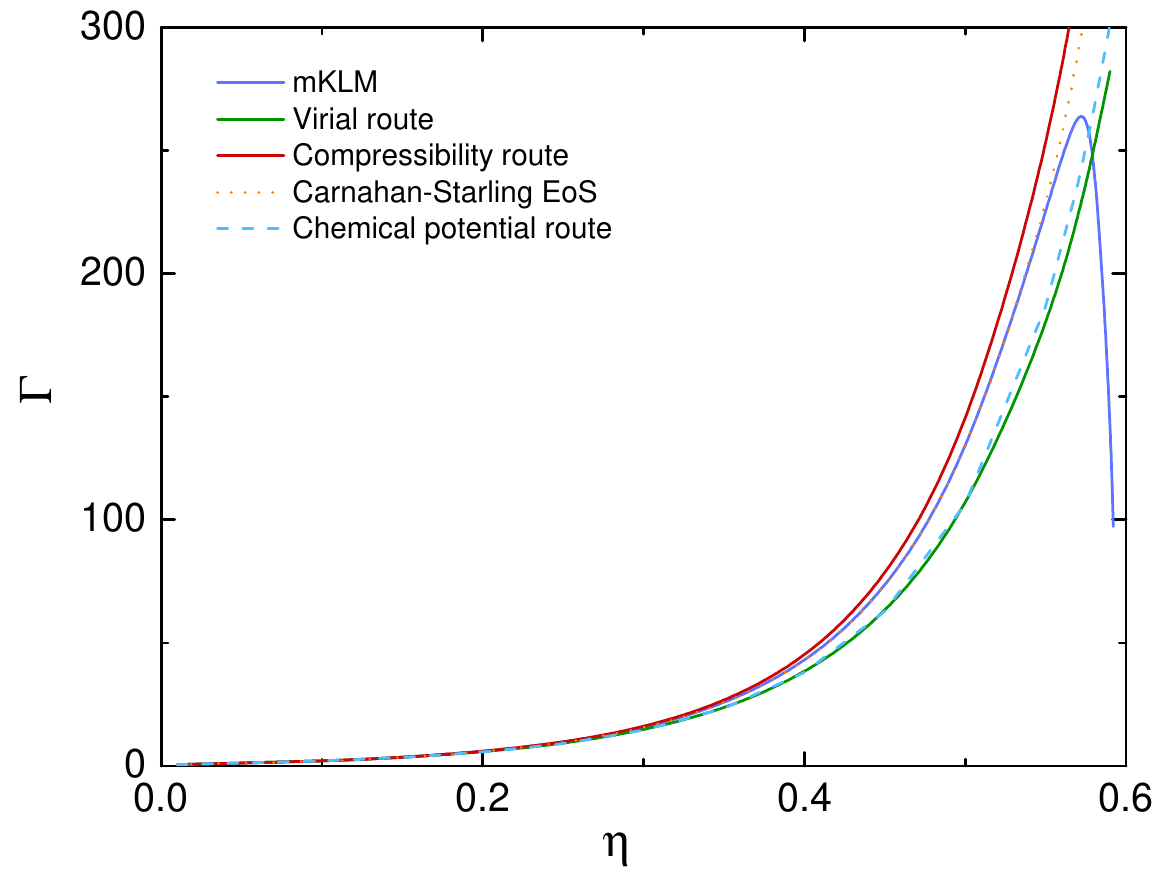}
\caption{(Color online) Dependence of $\Gamma$ on $\eta$ as calculated from Eq.~(\ref{Gamma}) using different HS compressibilities, see the legend.}
\label{Fig0}
\end{figure}

The dependence of $\Gamma$ on $\eta$ calculated from Eq.~(\ref{Gamma}) using different approximations for the HS EoS is shown in Fig.~\ref{Fig0}. The curves very nearly coincide at low values of $\eta$, but then start to diverge in the dense-fluid regime. In all approximations, except mKLM, $\Gamma$ grows monotonically with $\eta$. In contrast, in the mKLM approach, the coupling parameter demonstrates a nonphysical decrease with $\eta$ at high densities. However, this occurs at densities above the upper applicability limit of mKLM (at $\eta\gtrsim 0.55$) and therefore simply reflects the inaccuracy of mKLM EoS in this regime.

\begin{figure}
\includegraphics[width=8cm]{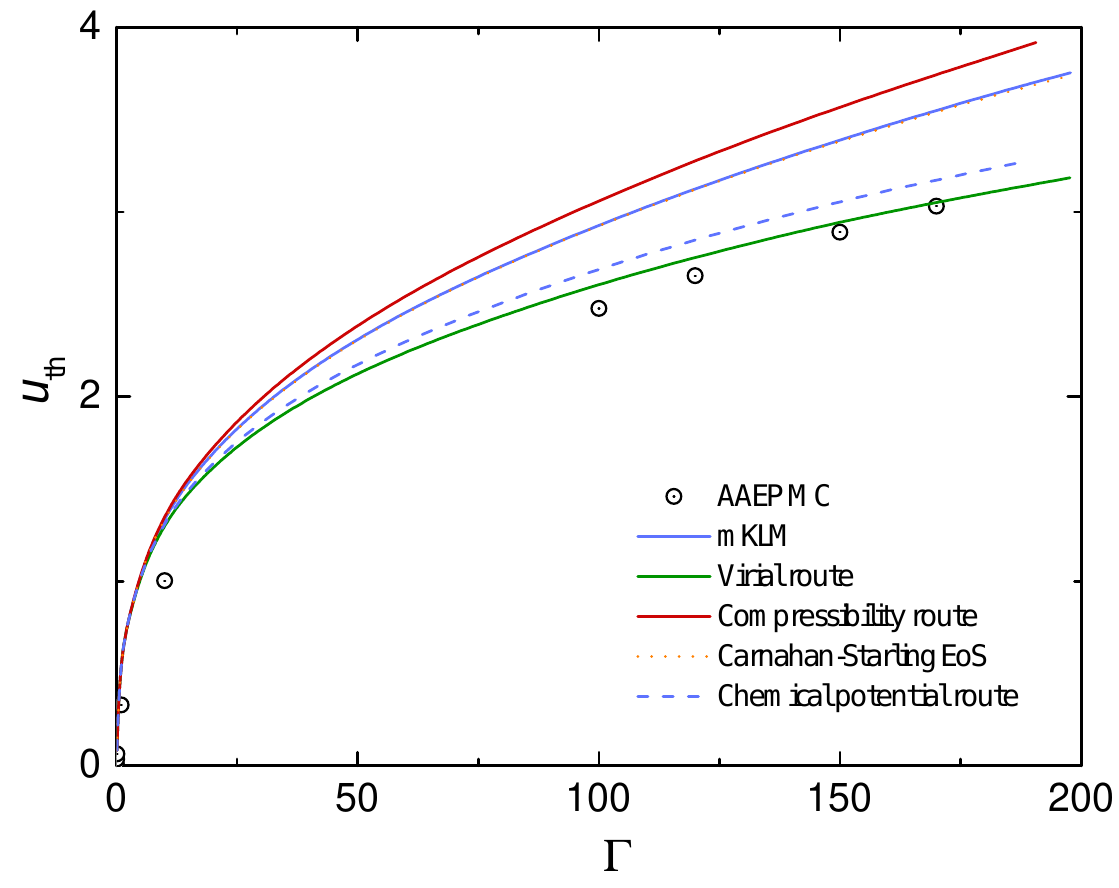}
\caption{(Color online) Thermal component of the excess energy of a strongly coupled OCP fluid. The circles correspond to AAEP MC numerical data from Ref.~\cite{DemyanovArxiv2025}. Different curves correspond to variational calculation using different HS entropy, see the legend. At strong coupling, PY theory supplemented by virial route provides the best agreement with MC data. }
\label{Fig1}
\end{figure}

Next, we calculate the excess energy of the OCP fluid using Eq.~(\ref{uex}). In order to make the comparison more evident, the fluid static energy (the fluid Madelung energy also known as the ion-sphere model energy~\cite{BausPR1980,IchimaruRMP1982,RosenfeldPRE2000,KhrapakPoP2014,KhrapakCPP2016}),
\begin{equation}\label{ISM}
u_{\rm st}=-\frac{9}{10}\Gamma,    
\end{equation}
has been subtracted from the total excess energy (Note that the result of the ion sphere model of Eq.~(\ref{ISM}) can be obtained by evaluating Eq.~(\ref{uex}) at the unphysical value of $\eta=1$; this is a general procedure that also works for other potentials; see, e.g.~\cite{KhrapakPoP2014} for the case of the Yukawa interaction potential). The remaining part is usually called the thermal component of the excess energy, since it depends on the temperature. It is plotted in Fig.~\ref{Fig1} along with recent MC data from Ref.~\cite{DemyanovArxiv2025}. Not unexpectedly, different HS EoS provide different levels of accuracy for the excess thermal energy of the OCP. What is perhaps surprising is that the most accurate mKLM and CS models do not reproduce the reference MC data well. The most accurate approximation is based on the PY virial route~\cite{DeWittPLA1979}. The approach based on the chemical potential route is slightly less accurate. Other approaches overestimate the thermal component of the excess energy.     

\section{Discussion and conclusion}

The first important observation is that the accuracy of the HS EoS and entropy {\it per se} does not guarantee the accuracy of the variational calculation for the excess energy of the OCP. Rather, the consistency between the RDF $g_0(r)$ used to evaluate the entropy of the HS system and the excess energy of the OCP system plays a much more important role. This was already mentioned by DeWitt and Rosenfeld~\cite{DeWittPLA1979} who noted that while the Carnahan-Starling EoS is better for hard spheres than either the PY virial or the PY compressibility EoS, the CS entropy is inconsistent with the PY $g_0(r)$ used to evaluate the excess energy of the OCP. However, the PY virial entropy of the HS fluid is consistent in the sense that it is obtained from the same $g_0(r)$ used to evaluate the excess energy of the OCP. This is why a more accurate mKLM EoS does not provide any improvement compared to other EoS; its performance in the variational calculation is very close to that of the CS EoS. The PY chemical potential route is conceptually closer to the PY virial route than to the PY compressibility route~\cite{SantosPRL2012}, and this can explain its better agreement with the MC results.    

The accuracy of the variational approach displayed in Fig.~\ref{Fig1} may not seem particularly exciting. However, it should be realized that the thermal component of the excess energy provides only a tiny fraction of the total excess energy for soft and long-range interactions such as in the OCP fluid. Near the fluid-solid phase transition, the thermal component contributes only about 2$\%$ of the total excess energy. This means that some inaccuracy in the thermal component will not significantly affect the exact value of the total excess energy. To give a concrete example, at $\Gamma=150$ the MC simulation result for $N=10^5$ particles is $u_{\rm ex}=-132.1104$~\cite{DemyanovArxiv2025}. The variational calculation that uses the PY virial HS EoS yields $u_{\rm ex}\simeq -132.06$, which corresponds to a relative deviation of only $0.04\%$. The largest deviation among the different HS EoS considered occurs for the PY compressibility route, but even in this case the calculated total excess energy is $u_{\rm ex}\simeq -131.41$, which corresponds to the relative deviation $0.5\%$. The remarkable accuracy of the variational approach with the PY virial HS entropy can potentially be useful in the context of other systems, such as, for instance, a strongly coupled Yukawa fluid.       

\begin{figure}
\includegraphics[width=8cm]{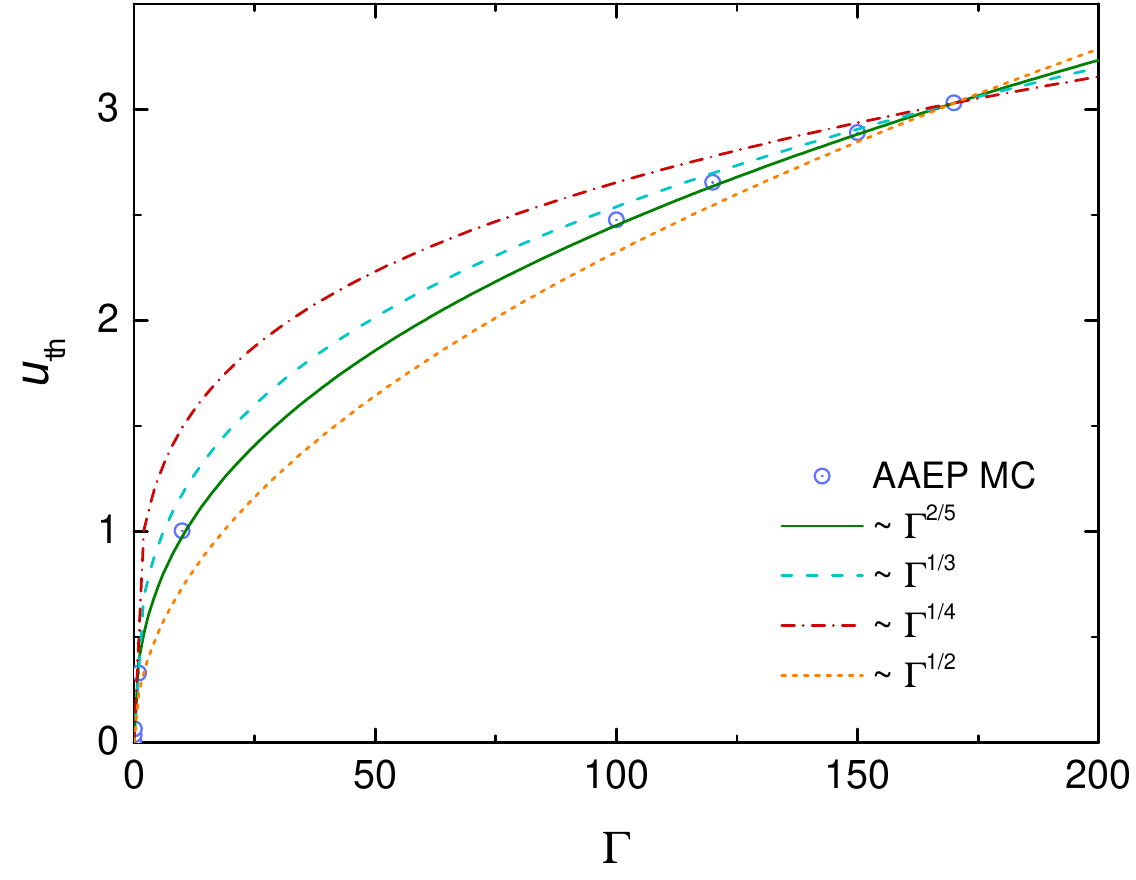}
\caption{(Color online) Thermal component of the excess energy of a strongly coupled OCP fluid. The circles correspond to AAEP MC numerical data from Ref.~\cite{DemyanovArxiv2025}. The curves correspond to the generalized RT scaling $u_{\rm th}\propto \Gamma^{\alpha}$ with different exponents $\alpha$. The original exponent $\alpha=2/5$ delivers better agreement with numerical results. }
\label{Fig2}
\end{figure}

Our final comment is related to the Rosenfeld-Tarazona scaling of the thermal component of the reduced excess energy.
Starting with a fundamental-measure hard-sphere reference functional, Rosenfeld and Tarazona (RT) were able
to demonstrate that the high density expansion for the potential energy of fluids is dominated by the fluid Madelung energy with a $\propto T^{3/5}$ thermal correction~\cite{RosenfeldMolPhys1998}. Using our current notation, their main result can be expressed as
\begin{equation}
u_{\rm th}(\Gamma)\simeq \alpha\left(\frac{\Gamma}{\Gamma_{\rm fr}}\right)^{2/5},
\end{equation}
where $\Gamma_{\rm fr}\simeq 174$ is the coupling parameter at the fluid-solid phase transition in the OCP, and $\alpha$ is a system-dependent parameter, often close to $\simeq 3$~\cite{RosenfeldPRE2000,KhrapakPRE02_2015,KhrapakJCP2015}. The original RT scaling works reasonably well for various systems~\cite{IngebrigtsenJCP2013} and represents a useful tool to develop simple practical EoS for systems with soft interactions~\cite{RosenfeldPRE2000,KhrapakPRE02_2015,KhrapakJCP2015}. Recently, using a large amount of available data on real and model systems, it has been empirically demonstrated that the original exponent $2/5$ of the RT scaling does not always represent the optimal choice~\cite{KhrapakPRE09_2024,KhrapakPOF11_2024}. Therefore, a generalized RT scaling has been proposed, in which the exponent is treated as a material-dependent parameter~\cite{KhrapakPOF11_2024}. Moreover, it can be shown that the generalized RT scaling emerges naturally in the two-phase model, which treats a liquid as a superposition of gas- and solid-like components whose relative abundance is quantified by a liquid rigidity parameter, from the scale invariance of this rigidity parameter~\cite{KhrapakJETPLett2025}. 

The present context represents an excellent opportunity to scrutinize the RT scaling in the special case of the OCP fluid.  The thermal component of the excess energy as obtained in recent AAEP MC simulation in Ref.~\cite{DemyanovArxiv2025} is plotted in Fig.~\ref{Fig2}. The various curves shown in the Figure correspond to the generalized RT scaling of the form $u_{\rm th}= \alpha(\Gamma/\Gamma_{\rm fr})^{\beta}$ with different exponents $\beta=1/4$, 1/3, 2/5 and 1/2. All curves have a common intersection point at $\Gamma = 170$ and $u_{\rm th}\simeq 3.03$. We observe that the best agreement with the MC results is provided by the original RT exponent $\beta=2/5$. The corresponding parameter $\alpha$ is $\alpha\simeq 3.06$. This correlates well with the value used to construct a simple practical EoS for the Yukawa fluid~\cite{RosenfeldPRE2000,KhrapakPRE02_2015,KhrapakJCP2015}.  

To conclude, we performed a variational calculation of the excess energy of the OCP fluid using different variants of the HS EoS, including three outcomes of the PY theory (using virial, compressibility, and chemical potential routes), Carnahan-Starling EoS, and perhaps the most accurate currently available mKLM EoS from Ref.~\cite{Pieprzyk2019}. Although the PY virial and CS approaches have previously been tested within the framework of the variational approach~\cite{StroudPRA1976,DeWittPLA1979}, the PY chemical potential route and mKLM EoS have not been used to the best of our knowledge. Our comparison demonstrates that the PY virial EoS agrees best with recent MC results for the excess energy of the OCP fluid. We have learned an important lesson: Consistency between the RDFs used to calculate the HS pressure and the OCP energy is more important than the accuracy of the HS EoS itself. A demonstrated very high accuracy at strong coupling gives hope that the approach can be useful for other related systems, in particular for the strongly coupled Yukawa fluid. This might merit verification in future work. We have also shown that the original RT scaling of the thermal component of the excess energy of the OCP fluid is in very good agreement with recent MC data.

The authors declare no conflict of interests.

The data that support the findings of this study are available from the authors upon reasonable request.

\bibliography{SE_Ref}

\providecommand{\noopsort}[1]{}\providecommand{\singleletter}[1]{#1}%
\begin{thebibliography}{40}%
\makeatletter
\providecommand \@ifxundefined [1]{%
 \@ifx{#1\undefined}
}%
\providecommand \@ifnum [1]{%
 \ifnum #1\expandafter \@firstoftwo
 \else \expandafter \@secondoftwo
 \fi
}%
\providecommand \@ifx [1]{%
 \ifx #1\expandafter \@firstoftwo
 \else \expandafter \@secondoftwo
 \fi
}%
\providecommand \natexlab [1]{#1}%
\providecommand \enquote  [1]{``#1''}%
\providecommand \bibnamefont  [1]{#1}%
\providecommand \bibfnamefont [1]{#1}%
\providecommand \citenamefont [1]{#1}%
\providecommand \href@noop [0]{\@secondoftwo}%
\providecommand \href [0]{\begingroup \@sanitize@url \@href}%
\providecommand \@href[1]{\@@startlink{#1}\@@href}%
\providecommand \@@href[1]{\endgroup#1\@@endlink}%
\providecommand \@sanitize@url [0]{\catcode `\\12\catcode `\$12\catcode
  `\&12\catcode `\#12\catcode `\^12\catcode `\_12\catcode `\%12\relax}%
\providecommand \@@startlink[1]{}%
\providecommand \@@endlink[0]{}%
\providecommand \url  [0]{\begingroup\@sanitize@url \@url }%
\providecommand \@url [1]{\endgroup\@href {#1}{\urlprefix }}%
\providecommand \urlprefix  [0]{URL }%
\providecommand \Eprint [0]{\href }%
\providecommand \doibase [0]{http://dx.doi.org/}%
\providecommand \selectlanguage [0]{\@gobble}%
\providecommand \bibinfo  [0]{\@secondoftwo}%
\providecommand \bibfield  [0]{\@secondoftwo}%
\providecommand \translation [1]{[#1]}%
\providecommand \BibitemOpen [0]{}%
\providecommand \bibitemStop [0]{}%
\providecommand \bibitemNoStop [0]{.\EOS\space}%
\providecommand \EOS [0]{\spacefactor3000\relax}%
\providecommand \BibitemShut  [1]{\csname bibitem#1\endcsname}%
\let\auto@bib@innerbib\@empty
\bibitem [{\citenamefont {Hansen}\ and\ \citenamefont
  {McDonald}(2006)}]{HansenBook}%
  \BibitemOpen
  \bibfield  {author} {\bibinfo {author} {\bibfnamefont {J.-P.}\ \bibnamefont
  {Hansen}}\ and\ \bibinfo {author} {\bibfnamefont {I.~R.}\ \bibnamefont
  {McDonald}},\ }\href@noop {} {\emph {\bibinfo {title} {Theory of Simple
  Liquids -}}}\ (\bibinfo  {publisher} {Elsevier},\ \bibinfo {address}
  {Amsterdam},\ \bibinfo {year} {2006})\BibitemShut {NoStop}%
\bibitem [{\citenamefont {Mansoori}\ and\ \citenamefont
  {Canfield}(1969{\natexlab{a}})}]{MansooriJCP1969a}%
  \BibitemOpen
  \bibfield  {author} {\bibinfo {author} {\bibfnamefont {G.~A.}\ \bibnamefont
  {Mansoori}}\ and\ \bibinfo {author} {\bibfnamefont {F.~B.}\ \bibnamefont
  {Canfield}},\ }\bibfield  {title} {\enquote {\bibinfo {title} {Variational
  approach to the equilibrium thermodynamic properties of simple liquids.
  {I}},}\ }\href {\doibase 10.1063/1.1671889} {\bibfield  {journal} {\bibinfo
  {journal} {J. Chem. Phys.}\ }\textbf {\bibinfo {volume} {51}},\ \bibinfo
  {pages} {4958–4967} (\bibinfo {year} {1969}{\natexlab{a}})}\BibitemShut
  {NoStop}%
\bibitem [{\citenamefont {Mansoori}\ and\ \citenamefont
  {Canfield}(1969{\natexlab{b}})}]{MansooriJCP1969b}%
  \BibitemOpen
  \bibfield  {author} {\bibinfo {author} {\bibfnamefont {G.~A.}\ \bibnamefont
  {Mansoori}}\ and\ \bibinfo {author} {\bibfnamefont {F.~B.}\ \bibnamefont
  {Canfield}},\ }\bibfield  {title} {\enquote {\bibinfo {title} {Variational
  approach to melting. {II}},}\ }\href {\doibase 10.1063/1.1671890} {\bibfield
  {journal} {\bibinfo  {journal} {J. Chem. Phys.}\ }\textbf {\bibinfo {volume}
  {51}},\ \bibinfo {pages} {4967–4972} (\bibinfo {year}
  {1969}{\natexlab{b}})}\BibitemShut {NoStop}%
\bibitem [{\citenamefont {Ross}(1973)}]{RossPRA1973}%
  \BibitemOpen
  \bibfield  {author} {\bibinfo {author} {\bibfnamefont {M.}~\bibnamefont
  {Ross}},\ }\bibfield  {title} {\enquote {\bibinfo {title} {Shock compression
  and the melting curve for argon},}\ }\href {\doibase 10.1103/physreva.8.1466}
  {\bibfield  {journal} {\bibinfo  {journal} {Phys. Rev. A}\ }\textbf {\bibinfo
  {volume} {8}},\ \bibinfo {pages} {1466–1474} (\bibinfo {year}
  {1973})}\BibitemShut {NoStop}%
\bibitem [{\citenamefont {Young}\ and\ \citenamefont
  {Rogers}(1984)}]{YoungJCP1984}%
  \BibitemOpen
  \bibfield  {author} {\bibinfo {author} {\bibfnamefont {D.~A.}\ \bibnamefont
  {Young}}\ and\ \bibinfo {author} {\bibfnamefont {F.~J.}\ \bibnamefont
  {Rogers}},\ }\bibfield  {title} {\enquote {\bibinfo {title} {Variational
  fluid theory with inverse 12th power reference potential},}\ }\href {\doibase
  10.1063/1.447951} {\bibfield  {journal} {\bibinfo  {journal} {J. Chem.
  Phys.}\ }\textbf {\bibinfo {volume} {81}},\ \bibinfo {pages} {2789–2793}
  (\bibinfo {year} {1984})}\BibitemShut {NoStop}%
\bibitem [{\citenamefont {Ross}\ \emph {et~al.}(1981)\citenamefont {Ross},
  \citenamefont {DeWitt},\ and\ \citenamefont {Hubbard}}]{RossPRA1981}%
  \BibitemOpen
  \bibfield  {author} {\bibinfo {author} {\bibfnamefont {M.}~\bibnamefont
  {Ross}}, \bibinfo {author} {\bibfnamefont {H.~E.}\ \bibnamefont {DeWitt}}, \
  and\ \bibinfo {author} {\bibfnamefont {W.~B.}\ \bibnamefont {Hubbard}},\
  }\bibfield  {title} {\enquote {\bibinfo {title} {{Monte Carlo} and
  perturbation-theory calculations for liquid metals},}\ }\href {\doibase
  10.1103/physreva.24.1016} {\bibfield  {journal} {\bibinfo  {journal} {Phys.
  Rev. A}\ }\textbf {\bibinfo {volume} {24}},\ \bibinfo {pages} {1016–1020}
  (\bibinfo {year} {1981})}\BibitemShut {NoStop}%
\bibitem [{\citenamefont {Baus}\ and\ \citenamefont
  {Hansen}(1980)}]{BausPR1980}%
  \BibitemOpen
  \bibfield  {author} {\bibinfo {author} {\bibfnamefont {M}~\bibnamefont
  {Baus}}\ and\ \bibinfo {author} {\bibfnamefont {J.~P.}\ \bibnamefont
  {Hansen}},\ }\bibfield  {title} {\enquote {\bibinfo {title} {Statistical
  mechanics of simple {C}oulomb systems},}\ }\href {\doibase
  10.1016/0370-1573(80)90022-8} {\bibfield  {journal} {\bibinfo  {journal}
  {Phys. Rep.}\ }\textbf {\bibinfo {volume} {59}},\ \bibinfo {pages} {1--94}
  (\bibinfo {year} {1980})}\BibitemShut {NoStop}%
\bibitem [{\citenamefont {Ichimaru}(1982)}]{IchimaruRMP1982}%
  \BibitemOpen
  \bibfield  {author} {\bibinfo {author} {\bibfnamefont {S.}~\bibnamefont
  {Ichimaru}},\ }\bibfield  {title} {\enquote {\bibinfo {title} {Strongly
  coupled plasmas: {H}igh-density classical plasmas and degenerate electron
  liquids},}\ }\href {\doibase 10.1103/revmodphys.54.1017} {\bibfield
  {journal} {\bibinfo  {journal} {Rev. Mod. Phys.}\ }\textbf {\bibinfo {volume}
  {54}},\ \bibinfo {pages} {1017--1059} (\bibinfo {year} {1982})}\BibitemShut
  {NoStop}%
\bibitem [{\citenamefont {Pieprzyk}\ \emph {et~al.}(2019)\citenamefont
  {Pieprzyk}, \citenamefont {Bannerman}, \citenamefont {Bra{\'{n}}ka},
  \citenamefont {Chudak},\ and\ \citenamefont {Heyes}}]{Pieprzyk2019}%
  \BibitemOpen
  \bibfield  {author} {\bibinfo {author} {\bibfnamefont {S.}~\bibnamefont
  {Pieprzyk}}, \bibinfo {author} {\bibfnamefont {M.~N.}\ \bibnamefont
  {Bannerman}}, \bibinfo {author} {\bibfnamefont {A.~C.}\ \bibnamefont
  {Bra{\'{n}}ka}}, \bibinfo {author} {\bibfnamefont {M.}~\bibnamefont
  {Chudak}}, \ and\ \bibinfo {author} {\bibfnamefont {D.~M.}\ \bibnamefont
  {Heyes}},\ }\bibfield  {title} {\enquote {\bibinfo {title} {Thermodynamic and
  dynamical properties of the hard sphere system revisited by molecular
  dynamics simulation},}\ }\href {\doibase 10.1039/c9cp00903e} {\bibfield
  {journal} {\bibinfo  {journal} {Phys. Chem. Chem. Phys.}\ }\textbf {\bibinfo
  {volume} {21}},\ \bibinfo {pages} {6886--6899} (\bibinfo {year}
  {2019})}\BibitemShut {NoStop}%
\bibitem [{\citenamefont {Demyanov}\ and\ \citenamefont
  {Levashov}(2022)}]{DemyanovPRE2022}%
  \BibitemOpen
  \bibfield  {author} {\bibinfo {author} {\bibfnamefont {G.~S.}\ \bibnamefont
  {Demyanov}}\ and\ \bibinfo {author} {\bibfnamefont {P.~R.}\ \bibnamefont
  {Levashov}},\ }\bibfield  {title} {\enquote {\bibinfo {title} {One-component
  plasma of a million particles via angular-averaged {E}wald potential: {A
  Monte Carlo} study},}\ }\href {\doibase 10.1103/physreve.106.015204}
  {\bibfield  {journal} {\bibinfo  {journal} {Phys. Rev. E}\ }\textbf {\bibinfo
  {volume} {106}},\ \bibinfo {pages} {015204} (\bibinfo {year}
  {2022})}\BibitemShut {NoStop}%
\bibitem [{\citenamefont {Demyanov}\ and\ \citenamefont {Levashov}(arXiv,
  2025)}]{DemyanovArxiv2025}%
  \BibitemOpen
  \bibfield  {author} {\bibinfo {author} {\bibfnamefont {G.~S.}\ \bibnamefont
  {Demyanov}}\ and\ \bibinfo {author} {\bibfnamefont {P.~R.}\ \bibnamefont
  {Levashov}},\ }\href {\doibase 10.48550/ARXIV.2509.02390} {\enquote {\bibinfo
  {title} {One--component plasma equation of state revisited via
  angular--averaged ewald potential},}\ } (\bibinfo {year} {arXiv,
  2025})\BibitemShut {NoStop}%
\bibitem [{\citenamefont {Huang}\ \emph {et~al.}(2023)\citenamefont {Huang},
  \citenamefont {Baggioli}, \citenamefont {Lu}, \citenamefont {Ma},\ and\
  \citenamefont {Feng}}]{HuangPRR2023}%
  \BibitemOpen
  \bibfield  {author} {\bibinfo {author} {\bibfnamefont {D.}~\bibnamefont
  {Huang}}, \bibinfo {author} {\bibfnamefont {M.}~\bibnamefont {Baggioli}},
  \bibinfo {author} {\bibfnamefont {S.}~\bibnamefont {Lu}}, \bibinfo {author}
  {\bibfnamefont {Z.}~\bibnamefont {Ma}}, \ and\ \bibinfo {author}
  {\bibfnamefont {Y.}~\bibnamefont {Feng}},\ }\bibfield  {title} {\enquote
  {\bibinfo {title} {Revealing the supercritical dynamics of dusty plasmas and
  their liquidlike to gaslike dynamical crossover},}\ }\href {\doibase
  10.1103/physrevresearch.5.013149} {\bibfield  {journal} {\bibinfo  {journal}
  {Phys. Rev. Research}\ }\textbf {\bibinfo {volume} {5}},\ \bibinfo {pages}
  {013149} (\bibinfo {year} {2023})}\BibitemShut {NoStop}%
\bibitem [{\citenamefont {Khrapak}\ and\ \citenamefont
  {Khrapak}(2022)}]{KhrapakPoF2022}%
  \BibitemOpen
  \bibfield  {author} {\bibinfo {author} {\bibfnamefont {S.~A.}\ \bibnamefont
  {Khrapak}}\ and\ \bibinfo {author} {\bibfnamefont {A.~G.}\ \bibnamefont
  {Khrapak}},\ }\bibfield  {title} {\enquote {\bibinfo {title} {Minima of shear
  viscosity and thermal conductivity coefficients of classical fluids},}\
  }\href {\doibase 10.1063/5.0082465} {\bibfield  {journal} {\bibinfo
  {journal} {Phys. Fluids}\ }\textbf {\bibinfo {volume} {34}},\ \bibinfo
  {pages} {027102} (\bibinfo {year} {2022})}\BibitemShut {NoStop}%
\bibitem [{\citenamefont {Khrapak}\ \emph {et~al.}(2025)\citenamefont
  {Khrapak}, \citenamefont {Wu},\ and\ \citenamefont {Du}}]{KhrapakPoP2025}%
  \BibitemOpen
  \bibfield  {author} {\bibinfo {author} {\bibfnamefont {S.~A.}\ \bibnamefont
  {Khrapak}}, \bibinfo {author} {\bibfnamefont {Y.-F.}\ \bibnamefont {Wu}}, \
  and\ \bibinfo {author} {\bibfnamefont {C.-R.}\ \bibnamefont {Du}},\
  }\bibfield  {title} {\enquote {\bibinfo {title} {A binary collision approach
  in one-component plasma: {H}ow close to the {F}renkel line can we go?}}\
  }\href {\doibase 10.1063/5.0287762} {\bibfield  {journal} {\bibinfo
  {journal} {Phys. Plasmas}\ }\textbf {\bibinfo {volume} {32}},\ \bibinfo
  {pages} {092114} (\bibinfo {year} {2025})}\BibitemShut {NoStop}%
\bibitem [{\citenamefont {Khrapak}(2021)}]{KhrapakMolecules12_2021}%
  \BibitemOpen
  \bibfield  {author} {\bibinfo {author} {\bibfnamefont {S.~A.}\ \bibnamefont
  {Khrapak}},\ }\bibfield  {title} {\enquote {\bibinfo {title} {Self-diffusion
  in simple liquids as a random walk process},}\ }\href {\doibase
  10.3390/molecules26247499} {\bibfield  {journal} {\bibinfo  {journal}
  {Molecules}\ }\textbf {\bibinfo {volume} {26}},\ \bibinfo {pages} {7499}
  (\bibinfo {year} {2021})}\BibitemShut {NoStop}%
\bibitem [{\citenamefont {Khrapak}(2024{\natexlab{a}})}]{KhrapakPhysRep2024}%
  \BibitemOpen
  \bibfield  {author} {\bibinfo {author} {\bibfnamefont {S.A.}\ \bibnamefont
  {Khrapak}},\ }\bibfield  {title} {\enquote {\bibinfo {title} {Elementary
  vibrational model for transport properties of dense fluids},}\ }\href
  {\doibase 10.1016/j.physrep.2023.11.004} {\bibfield  {journal} {\bibinfo
  {journal} {Phys. Rep.}\ }\textbf {\bibinfo {volume} {1050}},\ \bibinfo
  {pages} {1} (\bibinfo {year} {2024}{\natexlab{a}})}\BibitemShut {NoStop}%
\bibitem [{\citenamefont {Dubin}\ and\ \citenamefont
  {O'Neil}(1999)}]{DubinRMP1999}%
  \BibitemOpen
  \bibfield  {author} {\bibinfo {author} {\bibfnamefont {D.~H.~E.}\
  \bibnamefont {Dubin}}\ and\ \bibinfo {author} {\bibfnamefont {T.~M.}\
  \bibnamefont {O'Neil}},\ }\bibfield  {title} {\enquote {\bibinfo {title}
  {Trapped nonneutral plasmas, liquids, and crystals (the thermal equilibrium
  states)},}\ }\href {\doibase 10.1103/revmodphys.71.87} {\bibfield  {journal}
  {\bibinfo  {journal} {Rev. Mod. Phys.}\ }\textbf {\bibinfo {volume} {71}},\
  \bibinfo {pages} {87--172} (\bibinfo {year} {1999})}\BibitemShut {NoStop}%
\bibitem [{\citenamefont {Khrapak}\ and\ \citenamefont
  {Khrapak}(2016)}]{KhrapakCPP2016}%
  \BibitemOpen
  \bibfield  {author} {\bibinfo {author} {\bibfnamefont {S.~A.}\ \bibnamefont
  {Khrapak}}\ and\ \bibinfo {author} {\bibfnamefont {A.~G.}\ \bibnamefont
  {Khrapak}},\ }\bibfield  {title} {\enquote {\bibinfo {title} {Internal energy
  of the classical two- and three-dimensional one-component-plasma},}\ }\href
  {\doibase 10.1002/ctpp.201500104} {\bibfield  {journal} {\bibinfo  {journal}
  {Contrib. Plasma Phys.}\ }\textbf {\bibinfo {volume} {56}},\ \bibinfo {pages}
  {270--280} (\bibinfo {year} {2016})}\BibitemShut {NoStop}%
\bibitem [{\citenamefont {Tanaka}\ and\ \citenamefont
  {Ichimaru}(1987)}]{TanakaPRA1987}%
  \BibitemOpen
  \bibfield  {author} {\bibinfo {author} {\bibfnamefont {S.}~\bibnamefont
  {Tanaka}}\ and\ \bibinfo {author} {\bibfnamefont {S.}~\bibnamefont
  {Ichimaru}},\ }\bibfield  {title} {\enquote {\bibinfo {title} {Dynamic theory
  of correlations in strongly coupled, classical one-component plasmas: {G}lass
  transition in the generalized viscoelastic formalism},}\ }\href {\doibase
  10.1103/physreva.35.4743} {\bibfield  {journal} {\bibinfo  {journal} {Phys.
  Rev. A}\ }\textbf {\bibinfo {volume} {35}},\ \bibinfo {pages} {4743–4754}
  (\bibinfo {year} {1987})}\BibitemShut {NoStop}%
\bibitem [{\citenamefont {Ichimaru}\ and\ \citenamefont
  {Tanaka}(1986)}]{IchimaruPRL1986}%
  \BibitemOpen
  \bibfield  {author} {\bibinfo {author} {\bibfnamefont {S.}~\bibnamefont
  {Ichimaru}}\ and\ \bibinfo {author} {\bibfnamefont {S.}~\bibnamefont
  {Tanaka}},\ }\bibfield  {title} {\enquote {\bibinfo {title} {Generalized
  viscoelastic theory of the glass transition for strongly coupled, classical,
  one-component plasmas},}\ }\href {\doibase 10.1103/physrevlett.56.2815}
  {\bibfield  {journal} {\bibinfo  {journal} {Phys. Rev. Lett.}\ }\textbf
  {\bibinfo {volume} {56}},\ \bibinfo {pages} {2815} (\bibinfo {year}
  {1986})}\BibitemShut {NoStop}%
\bibitem [{\citenamefont {Cardenas}\ and\ \citenamefont
  {Tosi}(2004)}]{CardenasPhysB2004}%
  \BibitemOpen
  \bibfield  {author} {\bibinfo {author} {\bibfnamefont {M}~\bibnamefont
  {Cardenas}}\ and\ \bibinfo {author} {\bibfnamefont {M.P}\ \bibnamefont
  {Tosi}},\ }\bibfield  {title} {\enquote {\bibinfo {title} {Mean-field theory
  of the glass transition in the one-component classical plasma},}\ }\href
  {\doibase 10.1016/j.physb.2004.05.022} {\bibfield  {journal} {\bibinfo
  {journal} {Phys. B: Condens. Matter}\ }\textbf {\bibinfo {volume} {351}},\
  \bibinfo {pages} {137} (\bibinfo {year} {2004})}\BibitemShut {NoStop}%
\bibitem [{\citenamefont {Yazdi}\ \emph {et~al.}(2014)\citenamefont {Yazdi},
  \citenamefont {Ivlev}, \citenamefont {Khrapak}, \citenamefont {Thomas},
  \citenamefont {Morfill}, \citenamefont {L\"{o}wen}, \citenamefont {Wysocki},\
  and\ \citenamefont {Sperl}}]{YazdiPRE2014}%
  \BibitemOpen
  \bibfield  {author} {\bibinfo {author} {\bibfnamefont {A.}~\bibnamefont
  {Yazdi}}, \bibinfo {author} {\bibfnamefont {A.}~\bibnamefont {Ivlev}},
  \bibinfo {author} {\bibfnamefont {S.}~\bibnamefont {Khrapak}}, \bibinfo
  {author} {\bibfnamefont {H.}~\bibnamefont {Thomas}}, \bibinfo {author}
  {\bibfnamefont {G.~E.}\ \bibnamefont {Morfill}}, \bibinfo {author}
  {\bibfnamefont {H.}~\bibnamefont {L\"{o}wen}}, \bibinfo {author}
  {\bibfnamefont {A.}~\bibnamefont {Wysocki}}, \ and\ \bibinfo {author}
  {\bibfnamefont {M.}~\bibnamefont {Sperl}},\ }\bibfield  {title} {\enquote
  {\bibinfo {title} {Glass-transition properties of {Y}ukawa potentials: From
  charged point particles to hard spheres},}\ }\href {\doibase
  10.1103/PhysRevE.89.063105} {\bibfield  {journal} {\bibinfo  {journal} {Phys.
  Rev. E}\ }\textbf {\bibinfo {volume} {89}},\ \bibinfo {pages} {063105}
  (\bibinfo {year} {2014})}\BibitemShut {NoStop}%
\bibitem [{\citenamefont {Khrapak}(2024{\natexlab{b}})}]{KhrapakJCP2024_glass}%
  \BibitemOpen
  \bibfield  {author} {\bibinfo {author} {\bibfnamefont {S.~A.}\ \bibnamefont
  {Khrapak}},\ }\bibfield  {title} {\enquote {\bibinfo {title} {Shoving model
  and the glass transition in one-component plasma},}\ }\href {\doibase
  10.1063/5.0207393} {\bibfield  {journal} {\bibinfo  {journal} {J. Chem.
  Phys.}\ }\textbf {\bibinfo {volume} {160}},\ \bibinfo {pages} {166101}
  (\bibinfo {year} {2024}{\natexlab{b}})}\BibitemShut {NoStop}%
\bibitem [{\citenamefont {Wertheim}(1963)}]{Wertheim1963}%
  \BibitemOpen
  \bibfield  {author} {\bibinfo {author} {\bibfnamefont {M.~S.}\ \bibnamefont
  {Wertheim}},\ }\bibfield  {title} {\enquote {\bibinfo {title} {Exact solution
  of the {Percus-Yevick} integral equation for hard spheres},}\ }\href
  {\doibase 10.1103/physrevlett.10.321} {\bibfield  {journal} {\bibinfo
  {journal} {Phys. Rev. Lett.}\ }\textbf {\bibinfo {volume} {10}},\ \bibinfo
  {pages} {321--323} (\bibinfo {year} {1963})}\BibitemShut {NoStop}%
\bibitem [{\citenamefont {Jones}(1971)}]{JonesJCP1971}%
  \BibitemOpen
  \bibfield  {author} {\bibinfo {author} {\bibfnamefont {H.~D.}\ \bibnamefont
  {Jones}},\ }\bibfield  {title} {\enquote {\bibinfo {title} {Method for
  finding the equation of state of liquid metals},}\ }\href {\doibase
  10.1063/1.1676472} {\bibfield  {journal} {\bibinfo  {journal} {J. Chem.
  Phys.}\ }\textbf {\bibinfo {volume} {55}},\ \bibinfo {pages} {2640–2642}
  (\bibinfo {year} {1971})}\BibitemShut {NoStop}%
\bibitem [{\citenamefont {Stroud}\ and\ \citenamefont
  {Ashcroft}(1976)}]{StroudPRA1976}%
  \BibitemOpen
  \bibfield  {author} {\bibinfo {author} {\bibfnamefont {D.}~\bibnamefont
  {Stroud}}\ and\ \bibinfo {author} {\bibfnamefont {N.~W.}\ \bibnamefont
  {Ashcroft}},\ }\bibfield  {title} {\enquote {\bibinfo {title} {Comment on the
  thermodynamics of a classical one-component plasma},}\ }\href {\doibase
  10.1103/physreva.13.1660} {\bibfield  {journal} {\bibinfo  {journal} {Phys.
  Rev. A}\ }\textbf {\bibinfo {volume} {13}},\ \bibinfo {pages} {1660–1663}
  (\bibinfo {year} {1976})}\BibitemShut {NoStop}%
\bibitem [{\citenamefont {Thiele}(1963)}]{Thiele1963}%
  \BibitemOpen
  \bibfield  {author} {\bibinfo {author} {\bibfnamefont {E.}~\bibnamefont
  {Thiele}},\ }\bibfield  {title} {\enquote {\bibinfo {title} {Equation of
  state for hard spheres},}\ }\href {\doibase 10.1063/1.1734272} {\bibfield
  {journal} {\bibinfo  {journal} {J. Chem. Phys.}\ }\textbf {\bibinfo {volume}
  {39}},\ \bibinfo {pages} {474–479} (\bibinfo {year} {1963})}\BibitemShut
  {NoStop}%
\bibitem [{\citenamefont {Carnahan}\ and\ \citenamefont
  {Starling}(1969)}]{CarnahanJCP1969}%
  \BibitemOpen
  \bibfield  {author} {\bibinfo {author} {\bibfnamefont {N.~F.}\ \bibnamefont
  {Carnahan}}\ and\ \bibinfo {author} {\bibfnamefont {K.~E.}\ \bibnamefont
  {Starling}},\ }\bibfield  {title} {\enquote {\bibinfo {title} {Equation of
  state for nonattracting rigid spheres},}\ }\href {\doibase 10.1063/1.1672048}
  {\bibfield  {journal} {\bibinfo  {journal} {J. Chem. Phys.}\ }\textbf
  {\bibinfo {volume} {51}},\ \bibinfo {pages} {635--636} (\bibinfo {year}
  {1969})}\BibitemShut {NoStop}%
\bibitem [{\citenamefont {Santos}(2012)}]{SantosPRL2012}%
  \BibitemOpen
  \bibfield  {author} {\bibinfo {author} {\bibfnamefont {A.}~\bibnamefont
  {Santos}},\ }\bibfield  {title} {\enquote {\bibinfo {title}
  {Chemical-potential route: A hidden {Percus-Yevick} equation of state for
  hard spheres},}\ }\href {\doibase 10.1103/physrevlett.109.120601} {\bibfield
  {journal} {\bibinfo  {journal} {Phys. Rev. Lett.}\ }\textbf {\bibinfo
  {volume} {109}},\ \bibinfo {pages} {120601} (\bibinfo {year}
  {2012})}\BibitemShut {NoStop}%
\bibitem [{\citenamefont {Kolafa}\ \emph {et~al.}(2004)\citenamefont {Kolafa},
  \citenamefont {Labík},\ and\ \citenamefont {Malijevský}}]{KolafaPCCP2004}%
  \BibitemOpen
  \bibfield  {author} {\bibinfo {author} {\bibfnamefont {J.}~\bibnamefont
  {Kolafa}}, \bibinfo {author} {\bibfnamefont {S.}~\bibnamefont {Labík}}, \
  and\ \bibinfo {author} {\bibfnamefont {A.}~\bibnamefont {Malijevský}},\
  }\bibfield  {title} {\enquote {\bibinfo {title} {Accurate equation of state
  of the hard sphere fluid in stable and metastable regions},}\ }\href
  {\doibase 10.1039/b402792b} {\bibfield  {journal} {\bibinfo  {journal} {Phys.
  Chem. Chem. Phys.}\ }\textbf {\bibinfo {volume} {6}},\ \bibinfo {pages}
  {2335–2340} (\bibinfo {year} {2004})}\BibitemShut {NoStop}%
\bibitem [{\citenamefont {Rosenfeld}(2000)}]{RosenfeldPRE2000}%
  \BibitemOpen
  \bibfield  {author} {\bibinfo {author} {\bibfnamefont {Y.}~\bibnamefont
  {Rosenfeld}},\ }\bibfield  {title} {\enquote {\bibinfo {title}
  {Excess-entropy and freezing-temperature scalings for transport coefficients:
  Self-diffusion in {Y}ukawa systems},}\ }\href {\doibase
  10.1103/physreve.62.7524} {\bibfield  {journal} {\bibinfo  {journal} {Phys.
  Rev. E}\ }\textbf {\bibinfo {volume} {62}},\ \bibinfo {pages} {7524--7527}
  (\bibinfo {year} {2000})}\BibitemShut {NoStop}%
\bibitem [{\citenamefont {Khrapak}\ \emph {et~al.}(2014)\citenamefont
  {Khrapak}, \citenamefont {Khrapak}, \citenamefont {Ivlev},\ and\
  \citenamefont {Thomas}}]{KhrapakPoP2014}%
  \BibitemOpen
  \bibfield  {author} {\bibinfo {author} {\bibfnamefont {S.~A.}\ \bibnamefont
  {Khrapak}}, \bibinfo {author} {\bibfnamefont {A.~G.}\ \bibnamefont
  {Khrapak}}, \bibinfo {author} {\bibfnamefont {A.~V.}\ \bibnamefont {Ivlev}},
  \ and\ \bibinfo {author} {\bibfnamefont {H.~M.}\ \bibnamefont {Thomas}},\
  }\bibfield  {title} {\enquote {\bibinfo {title} {Ion sphere model for
  {Y}ukawa systems (dusty plasmas)},}\ }\href {\doibase 10.1063/1.4904309}
  {\bibfield  {journal} {\bibinfo  {journal} {Phys. Plasmas}\ }\textbf
  {\bibinfo {volume} {21}},\ \bibinfo {pages} {123705} (\bibinfo {year}
  {2014})}\BibitemShut {NoStop}%
\bibitem [{\citenamefont {DeWitt}\ and\ \citenamefont
  {Rosenfeld}(1979)}]{DeWittPLA1979}%
  \BibitemOpen
  \bibfield  {author} {\bibinfo {author} {\bibfnamefont {H.E.}\ \bibnamefont
  {DeWitt}}\ and\ \bibinfo {author} {\bibfnamefont {Y.}~\bibnamefont
  {Rosenfeld}},\ }\bibfield  {title} {\enquote {\bibinfo {title} {Derivation of
  the one component plasma fluid equation of state in strong coupling},}\
  }\href {\doibase 10.1016/0375-9601(79)90283-4} {\bibfield  {journal}
  {\bibinfo  {journal} {Phys. Lett. A}\ }\textbf {\bibinfo {volume} {75}},\
  \bibinfo {pages} {79–80} (\bibinfo {year} {1979})}\BibitemShut {NoStop}%
\bibitem [{\citenamefont {Rosenfeld}\ and\ \citenamefont
  {Tarazona}(1998)}]{RosenfeldMolPhys1998}%
  \BibitemOpen
  \bibfield  {author} {\bibinfo {author} {\bibfnamefont {Y.}~\bibnamefont
  {Rosenfeld}}\ and\ \bibinfo {author} {\bibfnamefont {P.}~\bibnamefont
  {Tarazona}},\ }\bibfield  {title} {\enquote {\bibinfo {title} {Density
  functional theory and the asymptotic high density expansion of the free
  energy of classical solids and fluids},}\ }\href {\doibase
  10.1080/00268979809483145} {\bibfield  {journal} {\bibinfo  {journal} {Mol.
  Phys.}\ }\textbf {\bibinfo {volume} {95}},\ \bibinfo {pages} {141--150}
  (\bibinfo {year} {1998})}\BibitemShut {NoStop}%
\bibitem [{\citenamefont {Khrapak}\ and\ \citenamefont
  {Thomas}(2015)}]{KhrapakPRE02_2015}%
  \BibitemOpen
  \bibfield  {author} {\bibinfo {author} {\bibfnamefont {S.~A.}\ \bibnamefont
  {Khrapak}}\ and\ \bibinfo {author} {\bibfnamefont {H.~M.}\ \bibnamefont
  {Thomas}},\ }\bibfield  {title} {\enquote {\bibinfo {title} {Practical
  expressions for the internal energy and pressure of {Y}ukawa fluids},}\
  }\href {\doibase 10.1103/physreve.91.023108} {\bibfield  {journal} {\bibinfo
  {journal} {Phys. Rev. E}\ }\textbf {\bibinfo {volume} {91}},\ \bibinfo
  {pages} {023108} (\bibinfo {year} {2015})}\BibitemShut {NoStop}%
\bibitem [{\citenamefont {Khrapak}\ \emph {et~al.}(2015)\citenamefont
  {Khrapak}, \citenamefont {Kryuchkov}, \citenamefont {Yurchenko},\ and\
  \citenamefont {Thomas}}]{KhrapakJCP2015}%
  \BibitemOpen
  \bibfield  {author} {\bibinfo {author} {\bibfnamefont {S.~A.}\ \bibnamefont
  {Khrapak}}, \bibinfo {author} {\bibfnamefont {N.~P.}\ \bibnamefont
  {Kryuchkov}}, \bibinfo {author} {\bibfnamefont {S.~O.}\ \bibnamefont
  {Yurchenko}}, \ and\ \bibinfo {author} {\bibfnamefont {H.~M.}\ \bibnamefont
  {Thomas}},\ }\bibfield  {title} {\enquote {\bibinfo {title} {Practical
  thermodynamics of {Y}ukawa systems at strong coupling},}\ }\href {\doibase
  10.1063/1.4921223} {\bibfield  {journal} {\bibinfo  {journal} {J. Chem.
  Phys.}\ }\textbf {\bibinfo {volume} {142}},\ \bibinfo {pages} {194903}
  (\bibinfo {year} {2015})}\BibitemShut {NoStop}%
\bibitem [{\citenamefont {Ingebrigtsen}\ \emph {et~al.}(2013)\citenamefont
  {Ingebrigtsen}, \citenamefont {Veldhorst}, \citenamefont {Schr{\o}der},\ and\
  \citenamefont {Dyre}}]{IngebrigtsenJCP2013}%
  \BibitemOpen
  \bibfield  {author} {\bibinfo {author} {\bibfnamefont {T.~S.}\ \bibnamefont
  {Ingebrigtsen}}, \bibinfo {author} {\bibfnamefont {A.~A.}\ \bibnamefont
  {Veldhorst}}, \bibinfo {author} {\bibfnamefont {T.~B.}\ \bibnamefont
  {Schr{\o}der}}, \ and\ \bibinfo {author} {\bibfnamefont {J.~C.}\ \bibnamefont
  {Dyre}},\ }\bibfield  {title} {\enquote {\bibinfo {title} {Communication: The
  {R}osenfeld-{T}arazona expression for liquids' specific heat: A numerical
  investigation of eighteen systems},}\ }\href {\doibase 10.1063/1.4827865}
  {\bibfield  {journal} {\bibinfo  {journal} {J. Chem. Phys.}\ }\textbf
  {\bibinfo {volume} {139}},\ \bibinfo {pages} {171101} (\bibinfo {year}
  {2013})}\BibitemShut {NoStop}%
\bibitem [{\citenamefont {Khrapak}(2024{\natexlab{c}})}]{KhrapakPRE09_2024}%
  \BibitemOpen
  \bibfield  {author} {\bibinfo {author} {\bibfnamefont {S.~A.}\ \bibnamefont
  {Khrapak}},\ }\bibfield  {title} {\enquote {\bibinfo {title} {Entropy of
  strongly coupled {Y}ukawa fluids},}\ }\href {\doibase
  10.1103/physreve.110.034602} {\bibfield  {journal} {\bibinfo  {journal}
  {Phys. Rev. E}\ }\textbf {\bibinfo {volume} {110}},\ \bibinfo {pages}
  {034602} (\bibinfo {year} {2024}{\natexlab{c}})}\BibitemShut {NoStop}%
\bibitem [{\citenamefont {Khrapak}\ and\ \citenamefont
  {Khrapak}(2024)}]{KhrapakPOF11_2024}%
  \BibitemOpen
  \bibfield  {author} {\bibinfo {author} {\bibfnamefont {S.~A.}\ \bibnamefont
  {Khrapak}}\ and\ \bibinfo {author} {\bibfnamefont {A.~G.}\ \bibnamefont
  {Khrapak}},\ }\bibfield  {title} {\enquote {\bibinfo {title} {Generalized
  {R}osenfeld–{T}arazona scaling and high-density specific heat of simple
  liquids},}\ }\href {\doibase 10.1063/5.0230219} {\bibfield  {journal}
  {\bibinfo  {journal} {Phys. Fluids}\ }\textbf {\bibinfo {volume} {36}},\
  \bibinfo {pages} {117119} (\bibinfo {year} {2024})}\BibitemShut {NoStop}%
\bibitem [{\citenamefont {Khrapak}\ and\ \citenamefont
  {Khrapak}(2025)}]{KhrapakJETPLett2025}%
  \BibitemOpen
  \bibfield  {author} {\bibinfo {author} {\bibfnamefont {S.~A.}\ \bibnamefont
  {Khrapak}}\ and\ \bibinfo {author} {\bibfnamefont {A.~G.}\ \bibnamefont
  {Khrapak}},\ }\bibfield  {title} {\enquote {\bibinfo {title} {Heat capacity
  of dense liquids: {A} link between two-phase model and melting temperature
  scaling},}\ }\href {\doibase 10.1134/s0021364025607262} {\bibfield  {journal}
  {\bibinfo  {journal} {JETP Lett.}\ }\textbf {\bibinfo {volume} {122}},\
  \bibinfo {pages} {240–243} (\bibinfo {year} {2025})}\BibitemShut {NoStop}%
\end{thebibliography}%

\end{document}